\documentclass[preprint,12pt]{elsarticle}

\usepackage{graphicx}
\usepackage{amssymb}

\usepackage{lineno}

\usepackage{hyperref}

\usepackage{multirow, enumerate}
\usepackage{verbatim}

\usepackage[caption=false,font=footnotesize]{subfig}

\journal{Journal Name}

\newcommand{\pkg}[1] {\textbf{#1}}
\newcommand{\proglang}[1] {\textit{#1}}
\newcommand{\code}[1] {\texttt{#1}}

\begin{document}

\begin{frontmatter}


\title{\pkg{AutoWIG}: Automatic Generation of \proglang{Python} Bindings for \proglang{C++} Libraries}



\author[inria]{Pierre Fernique}
\address[inria]{Inria, EPI VirtualPlants, Montpellier, France}
\author[cirad,inria]{Christophe Pradal}
\address[cirad]{CIRAD, UMR AGAP, Montpellier, France}


\begin{abstract}
Most of \proglang{Python} and \proglang{R} scientific packages incorporate compiled scientific libraries to speed up the code and reuse legacy libraries.
While several semi-automatic solutions exist to wrap these compiled libraries, the process of wrapping a large library is cumbersome and time consuming.
In this paper, we introduce \pkg{AutoWIG}, a \proglang{Python} package that wraps automatically compiled libraries into high-level languages using \pkg{LLVM}/\pkg{Clang} technologies and the \pkg{Mako} templating engine.
Our approach is automatic, extensible, and applies to complex \proglang{C++} libraries, composed of thousands of classes or incorporating modern meta-programming constructs.
\end{abstract}

\begin{keyword}
\proglang{C++} \sep \proglang{Python} \sep automatic bindings generation


\end{keyword}

\end{frontmatter}












\section{Introduction}
Many scientific libraries are written in low-level programming languages such as \proglang{Fortran}, \proglang{C} and \proglang{C++}.
Such libraries entail the usage of the traditional edit/compile/execute cycle in order to produce high-performance programs.
This leads to low computer's processing time at the cost of high scientist's coding time.
At the opposite, scripting languages such as \proglang{Matlab}, \proglang{Octave} \citep[for numerical work]{EBHW14}, \proglang{Sage} \citep[for symbolic mathematics]{sage15}, \proglang{R} \citep[for statistical analyses]{R14} or \proglang{Python} \citep[for general purposes]{Oli07} provide an interactive framework that allows data scientists to explore their data, test new ideas, combine algorithmic approaches and evaluate their results on the fly.
However, code executed in these high-level languages tends to be slower that their compiled counterpart.
Due to growing interest into data science combined with hardware improvements in the last decades, such high-level programming languages have become very popular in various scientific fields.
Nevertheless, to overcome performance bottleneck in these languages, most scientific packages of scripting languages incorporate compiled libraries available within the scripting language interpreter.
For instance, \pkg{SciPy} \citep{JOP14}, a library for scientific computing in \proglang{Python}, is mainly based on routines implemented in \proglang{Fortran}, \proglang{C} and \proglang{C++}.
To access compiled code from an interpreter, a programmer has to write a collection of special wrapper functions (aka wrappers).
The role of these functions is to convert arguments and return values between the data representation in each language.
Although it is affordable for a library to write a few wrappers, the task becomes tedious if the library contains a large number of functions.
Moreover, the task is considerably more complex and time consuming if a library uses more advanced programming features such as pointers, arrays, classes, inheritance, templates, operators and overloaded functions.
\pkg{Cython} \citep{BBCD+11}, \pkg{Boost.Python} \citep{AG03}, \pkg{SWIG} \citep{Bea03}, \pkg{Rcpp} \citep{EFAC+11} and \pkg{F2PY} \citep{Pet09} are considered as classical approaches for wrapping \proglang{C}, \proglang{C++} and \proglang{Fortran} libraries to \proglang{Python}, \proglang{R} or other scripting languages but can only be considered as semi-automatic.
In fact, while these approaches certainly ease the way of generating wrappers, the process of writing and maintaining wrappers for large libraries is still cumbersome, time consuming and not really designed for evolving libraries.
Every change in the library interface implies a change in the wrapper code. Thus, developers have to synchronize two code bases that do not rely on the same kind of knowledge (i.e., \proglang{C++} vs wrapper definition).
To solve this issue, we provide an automatic approach for wrapping \proglang{C++} libraries.
The critical bottleneck in the construction of an automatic approach for wrapping compiled languages libraries is the need to perform the syntactic analysis of the input code, known as parsing.
Once the code has been parsed, it is possible to analyze its result for code introspection.
Code introspection is the ability to examine code components to know what they represent and what are their relations to other code components (e.g., list all methods for a given class).
Introspection of parsed code can therefore be used to automate the generation of wrappers.

In the past, some solutions have been developed to automate the wrapping in \proglang{Python} of large  \proglang{C++} libraries such as \pkg{Py++} \citep{Yak11} and \pkg{XDress} \citep{Sco13}.
These tools require to write \emph{a priori} complex scripts.
These scripts are then interpreted \emph{a posteriori} to edit the code abstraction and generate wrappers.
Such batch processing approaches require high-level of expertise in these software and limit the ability to supervise or debug the wrapping process.
The cost of the wrapping processes with such methodologies, although automatic, is thus considered by many developers as prohibitive.
The goal of \pkg{AutoWIG} is to overcome these shortcomings.
\pkg{AutoWIG} proposes an interactive approach for the wrapping process and an extensible interface to easily incorporate bindings for other languages.
In particular, the proposed \proglang{Python} interface provides an easy-to-use environment in which the user can benefit of code introspection on large libraries.
The end-user can therefore analyze compiled library components, tests different wrapping strategies and evaluates their outcomes directly.

This paper is organized as follows.
Section~\ref{sec:requirements} provides an insight of requirements for an automated wrapping of compiled libraries.
Section~\ref{sec:workflow} presents the wrapping strategies that can be considered.
Section~\ref{sec:architecture} describes the main aspects of \pkg{AutoWIG}'s architecture and current implementations.
Section~\ref{sec:guidelines} presents \proglang{C++} coding guidelines that must be respected in order to obtain the most automated wrapping workflow.
Section~\ref{sec:results} presents different results of \pkg{AutoWIG} application including in particular examples for performing partial wrapping of a library, the wrapping of template libraries and the wrapping of dependent libraries using an actual \proglang{C++} statistical library set case study.
Note that, for the sake of simplicity in the remainder of this paper, it is assumed that the low-level programming language in which compiled libraries are written is \proglang{C++}, the high-level programming language for interfacing libraries is \proglang{Python} and, that the wrappers generated are written using the \pkg{Boost.Python} \proglang{C++} library.
Section~\ref{sec:discussion} will therefore be the occasion to discuss \pkg{AutoWIG}'s extensibility or limitations considering other programming languages.

\section{Requirements}
\label{sec:requirements}
Consider a scientist who has designed multiple \proglang{C++} libraries for statistical analysis.
He would like to distribute his libraries and decides to make them available in \proglang{Python} in order to reach a public of statisticians but also less expert scientists such as biologists.
Yet, he is not interested in becoming an expert in \proglang{C++}/\proglang{Python} wrapping, even if it exists classical approaches consisting in writing wrappers with \pkg{SWIG} or \pkg{Boost.Python}.
Moreover, he would have serious difficulties to maintain the wrappers, since this semi-automatic process is time consuming and error prone.
Instead, he would like to automate the process of generating wrappers in sync with his evolving \proglang{C++} libraries.
That's what the \pkg{AutoWIG} software aspires to achieve.
Building such a system entails achieving some minimal features:

\begin{description}
	\item[\proglang{C++} parsing] In order to automatically expose \proglang{C++} components in \proglang{Python}, the system requires parsing full legacy code implementing the last \proglang{C++} standard. It has also to represent \proglang{C++} constructs in \proglang{Python}, like namespaces, enumerators, enumerations, variables, functions, classes or aliases.
	\item[Pythonic interface]
          To respect the \proglang{Python} philosophy,  \proglang{C++} language patterns need to be consistently translated into \proglang{Python}.
    	  Some syntax or design patterns in \proglang{C++} code are specific and need to be adapted in order to obtain a functional \proglang{Python} package.
          Note that this is particularly sensible for \proglang{C++} operators (e.g., \code{()}, \code{<}, \code{[]}) and corresponding \proglang{Python} special functions (e.g., \code{\_\_call\_\_}, \code{\_\_lt\_\_}, \code{\_\_getitem\_\_}, \code{\_\_setitem\_\_}).
    \item[Memory management]
    	  \proglang{C++} libraries expose in their interfaces either raw pointers, shared pointers or references, while \proglang{Python} handles memory allocation and garbage collection automatically.
          The concepts of pointer and reference are thus not meaningful in \proglang{Python}.
          These language differences entail several problems in the memory management of \proglang{C++} components into \proglang{Python}.
          A special attention is therefore required for dealing with references (\code{\&}) and pointers (\code{*}) that are highly used in \proglang{C++}.
    \item[Error management]
          \proglang{C++} exceptions need to be consistently managed in \proglang{Python}.
          \proglang{Python} does not have the necessary equipment to properly unwind the \proglang{C++} stack when exceptions are thrown.
    	  It is therefore important to ensure that exceptions thrown by \proglang{C++} libraries do not pass into the \proglang{Python} interpreter core.
          All \proglang{C++} exceptions thrown by wrappers must therefore be translated into \proglang{Python} errors.
          Moreover, this translation must preserve the name and content of the exception in order to raise an informative \proglang{Python} error.
    \item[Dependency management between components] The management of multiple dependencies between \proglang{C++} libraries with \proglang{Python} bindings is required at run-time from \proglang{Python}.
          \proglang{C++} libraries tends to have dependencies.
          For instance the \proglang{C++} \pkg{Standard Template Library} containers \citep{PLMS00} are used in many \proglang{C++} libraries (e.g \code{std::vector}, \code{std::set}).
          For such cases, it does not seem relevant that every wrapped \proglang{C++} library contains wrappers for usual \pkg{STL} containers (e.g., \code{std::vector< double >}, \code{std::set< int >}).
          Moreover, loading in the \proglang{Python} interpreter multiple compiled libraries sharing different wrappers from same \proglang{C++} components could lead to serious side effects.
          It is therefore required that dependencies across different library bindings can be handled automatically.
    \item[Documentation]
          The documentation of \proglang{C++} components has to be associated automatically to their corresponding \proglang{Python} components in order to reduce the redundancy and to keep it up-to-date.
\end{description}

\section{Methodology}
\label{sec:workflow}



A major functionality of \pkg{AutoWIG} is its interactivity.
Interactive processing have some advantages versus batch processing.
In our context, such advantages are that an interactive framework allows developers to look at the abstraction of their code, to test new wrapping strategies and to evaluate their outcomes directly.
In such cases, the user must consider the following $3$ steps:
\begin{description}
	\item[\code{Parse}]
    	 In \proglang{C++} library, headers contain all declarations of usable \proglang{C++} components.
         This step performs a syntactic and a semantic analysis of these headers to obtain a proper abstraction of available \proglang{C++} components (see Section~\ref{subsec:architecture:plugin} for details).
    	 This abstraction is a graph database within which each \proglang{C++} component (namespaces, enumerators, enumerations, variables, functions, classes and aliases) used in the library are represented by a node.
         Edges connecting nodes in this graph database represent syntactic or semantic relation between nodes (see Section~\ref{subsec:architecture:asg} for details).
         Mandatory inputs of this workflow are headers and relevant compilation flags to conduct the \proglang{C++} code parsing (see Section~\ref{subsec:results:simple} for an example).
    \item[\code{Control}]
         Once the \code{Parse} step has been executed, the graph database can be used to interactively introspect the \proglang{C++} code.
         This step is particularly useful for controlling the output of the workflow.
         By default, \pkg{AutoWIG} has a set of rules for determining which \proglang{C++} components to wrap, selecting the adapted memory management, identifying special classes representing exceptions or smart pointers and adapting \proglang{C++} philosophy to \proglang{Python} (see Section~\ref{subsec:architecture:plugin} for details).
         Such rules produce consistent wrapping of \proglang{C++} libraries following precise guidelines (see Section~\ref{sec:guidelines} for details).
         This step enables the control of parameters to ensure a consistent wrapping of a \proglang{C++} library, even if it does not fully respect \pkg{AutoWIG} guidelines (see Section~\ref{subsec:results:clanglite} for an example).
    \item[\code{Generate}]
         Once control parameters have been correctly set in the \code{Control} step, the next step consists in the generation of wrapper functions for each \proglang{C++} component.
         This is also coupled with the generation of a pythonic interface for the \proglang{Python} module containing the wrappers (see Section~\ref{subsec:architecture:plugin} for details).
         This code generation step is based on graph database traversals and rules using \proglang{C++} code introspection realizable via the graph database (e.g., parent scope, type of variables, inputs and output of functions, class bases and members).
         The outputs of the workflow consists in \proglang{C++} files containing wrappers that need to be compiled and a \proglang{Python} file containing a pythonic interface for the \proglang{C++} library (see Section~\ref{subsec:results:simple} for an example).
\end{description}


If an interactive workflow is very convenient for first approaches with \pkg{AutoWIG}, once the wrapping strategies have been chosen, batch mode workflows are of great interest.
Note that the usage of the \pkg{IPython} console \citep{PG07} and its \code{\%history} magic function enable to save an interactive workflow into a \proglang{Python} file that can be executed in batch mode using the \code{python} command line.

\section{Architecture and implementation}
\label{sec:architecture}

In this section, we present the  architecture of \pkg{AutoWIG}, describe the technical design underlying the concepts introduced in Section~\ref{sec:workflow}, and discuss in details the implementation choices.
This section can be considered as technical and readers willing to focus first on \pkg{AutoWIG} big picture can jump to Section~\ref{sec:guidelines}.

\subsection{Data model}
\label{subsec:architecture:asg}

The central data model used in \pkg{AutoWIG} is an abstract semantic graph (ASG) that represent code abstraction and capture code components and their relationships.
In computer science, an ASG is a form of abstract syntax in which an expression of a programming language is represented by a graph whose nodes are its components.
This ASG principally contains nodes identified as file-system components (e.g., directories, files) or \proglang{C++} components (e.g., fundamental types, variables, functions, classes, aliases).
Syntactic and semantic relation between nodes are encoded either in edges (e.g., underlying type, inherited classes), edge properties (e.g., type qualifiers, base access) or node properties (e.g., method \code{static} or \code{const} qualifications, polymorphism of a class).


\subsection{Plugin architecture}
\label{subsec:architecture:plugin}

The software architecture is based on the concept of plugin (i.e., a component with a well-defined interface, that can be found dynamically and replaced by another one with the same interface).
Implementations can therefore be provided by the system or from a third-party.
Plugin architectures are attractive solutions for developers seeking to build applications that are modular, adaptive, and easily extensible.
A plugin manager (PM) is a component in charge of discovering and loading plugins that adhere to a specific contract.
As stated above, the wrapping process is decomposed into $3$ steps.
Each step is governed by a specific PM:
\begin{itemize}
	\item The \code{parser} PM is in charge of the \code{Parse} step.
          A \code{parser} plugin implements syntactic and semantic analyses of code in order to complete an existing ASG.
          Its inputs are an ASG (denoted \code{asg}), a set of source code files (denoted \code{headers}), compilation flags (denoted \code{flags}) and optional parameters (denoted \code{kwargs}).
          It returns a modified ASG.
    \item The \code{controller} PM is in charge of the \code{Control} step.
          A \code{controller} plugin enables workflow control.
          It ensures that code generated in the \code{Generate} step is flawless (e.g., ensure relevant memory management, hide undefined symbols or erroneous methods of class template specializations).
          Its inputs are an ASG and optional named parameters.
          It returns a modified ASG.
    \item The \code{generator} PM is in charge of the \code{Generate} step.
          A \code{generator} plugin interprets a node subset from the ASG for code generation.
          Its inputs are an ASG and optional parameters.
          It returns in-memory files (denoted \code{wrappers}) whose content corresponds to the generated code.
\end{itemize}

Considering these PMs, the workflow simply consists in passing the ASG step by step.
Plugin implementation requires different levels of expertise (see Table~\ref{table:architecture:scons}).
\begin{table}
	\begin{center}
    	\begin{tabular}{c|c|c|l}
        	Workflow & \multicolumn{1}{c}{ } & \multicolumn{1}{c}{Plugin} &\\
            step & manager & implementation & finality \\\hline\hline
            \multirow{4}{*}{\code{Parse}} & \multirow{4}{*}{\code{parser}} & \multirow{4}{*}{developer} & \multirow{4}{*}{\parbox{.4\textwidth}{Performs syntactic and semantic analysis of input code and produces an abstract semantic graph.}}\\
            & & & \\
            & & & \\
            & & & \\\hline
            \multirow{3}{*}{\code{Control}} & \multirow{3}{*}{\code{controller}} & \multirow{3}{*}{end-user} & \multirow{3}{*}{\parbox{.4\textwidth}{Regroups \proglang{Python} code editing the abstract semantic graph for workflow control.}}\\
            & & & \\
            & & & \\\hline
            \multirow{3}{*}{\code{Generate}} & \multirow{3}{*}{\code{generator}} & \multirow{3}{*}{developer} & \multirow{3}{*}{\parbox{.4\textwidth}{Traverses the abstract semantic graph and generates code given code generation rules.}}\\
            & & & \\
            & & & \\\hline
        \end{tabular}
        \caption{Plugin architecture of \pkg{AutoWIG}.
        Each step of the \pkg{AutoWIG} wrapping workflow is managed by a plugin manager that enables an easy control of the workflow outputs.
        Considering the finality and underlying complexity of these plugins, implementations responsibilities are shared between \pkg{AutoWIG} developers and end-users.
        The \code{parser} and \code{generator} plugins are respectively concerned with compiled and scripting languages admissible bindings.
        Since such implementations require a high-level of expertise and a variety of tests, they mostly concern \pkg{AutoWIG} developers.
        On the contrary, \code{controller} plugins are library dependent and only require the manipulation of the abstract semantic graph via \proglang{Python} code.
        Thus, most of \pkg{AutoWIG} end-users are concerned with \code{controller} implementations.}
        \label{table:architecture:scons}
    \end{center}
\end{table}
However, the registration of a new plugin in AutoWIG is simple due to the usage of the entry points mechanism provided by the \pkg{Setuptools} \proglang{Python} package.
Moreover, the concept of \pkg{AutoWIG} plugin manager enables an easy control of plugin implementation (see Section~\ref{subsec:results:pystl} for an example).

\paragraph{Parsers}
Currently, \pkg{AutoWIG} provides one \code{parser} for \proglang{C++} libraries.
Parsing \proglang{C++} is very challenging and mainly solved by compiler front-ends \citep{Gun11} that generate abstract syntax trees (ASTs).
There are many benefits in using a compiler front-end for parsing \proglang{C++} code.
In particular, the \code{parser} implementation simply uses the compiler front-end for performing syntactic and semantic analyses of code rather than performing itself a custom analysis of an evolving and complex language.
Therefore, the implementation mainly consists in AST traversals to complete ASGs, which is a far less challenging problem.
Since the development of \pkg{LLVM} \citep{LA04} and \pkg{Clang} \citep{Lat08} technologies, the AST, used for the compilation process, is directly available in \proglang{Python} via the \pkg{libclang} \proglang{Python} package.
Our \code{libclang} \code{parser} was therefore designed using \pkg{libclang}:
\begin{verbatim}
def libclang_parser(asg, headers, flags, bootstrap=True, **kwargs):
    header = pre_processing(asg, headers, flags, **kwargs)
    asg = processing(asg, header, flags, **kwargs)
    asg = post_processing(asg, flags, **kwargs)
    return asg
\end{verbatim}
This implementation consists in the three following steps:
\begin{description}
	\item[\code{Pre-process}]
          During the \code{pre\_processing} step, header files (\code{headers}) are added in the ASG and marked as self-contained headers (see Section~\ref{sec:guidelines} for details).
          Note that in order to distinguish headers of the current library from headers of external libraries that are included by these headers, the headers of the library are marked as internal dependency headers (opposed to external dependency headers).
          This step returns a temporary header (\code{header}) that includes all given headers.
          This approach enables to parse only one header including all others and therefore prevents the multiple and redundant parsing of headers.
          Note that compilation flags (\code{flags}) are also parsed in order to save \proglang{C++} search paths  (given by the \code{-I} option).
	\item[\code{Process}]
          During the \code{processing} step, the actual \proglang{C++} code is parsed using the \pkg{libclang} \proglang{Python} package.
          The parsing of the temporary header (\code{header}) returns an AST.
          The ASG is updated from the AST by a process of enrichment and abstraction.
          The enrichment entails the addition of node properties (e.g., if a class can be instantiated or copied, if a method is overloaded) or edges (e.g., forward-declarations, back-pointers to base classes, type of variables).
          The abstraction entails the removal of details which are relevant only in parsing, not for semantics (e.g., multiple opening and closing of namespaces).
	\item[\code{Post-process}]
          During the \code{post\_processing} step, the \proglang{C++} code is bootstrapped.
          Template class specializations are sometimes only declared but not defined (e.g., a template class specialization only used as a return type of a method).
          In order to have access to all template class specialization definitions, a virtual program in which definition of undefined template class specializations are ensured (e.g., using \code{sizeof(std::vector< int >);}  for forcing \code{std::vector< int >} definition) is parsed.
          Note that this step induces new undefined template class specializations and must therefore be repeated until no more undefined template class specializations arise.
          This step is controlled by the \code{bootstrap} parameter that can be set to \code{True}, \code{False} or an integer corresponding to the maximal number of repetition of this operation (\code{True} is equivalent to \code{bootstrap=float("inf")} and \code{False} to \code{bootstrap=0}).
\end{description}

\paragraph{Controllers}
By default, \pkg{AutoWIG} provides a \code{controller} for libraries respecting some recommended guidelines (see Section~\ref{sec:guidelines} for details):
\begin{verbatim}
def default_controller(asg, clean=True, **kwargs):
    asg = refactoring(asg, **kwargs)
    if clean:
        asg = cleaning(asg)
    return asg
\end{verbatim}
This \code{default} implementation consists of the two following steps:
\begin{description}
	\item[\code{Refactoring}]  The \code{refactoring} of the \proglang{C++} code is simulated in order to have  a wrapping compliant with \proglang{Python} rules.
         In \proglang{C++}, some operators (e.g., \code{operator+}) can be defined at the class scope or at the global scope.
         But in \proglang{Python}, special methods corresponding to these operators (e.g., \code{\_\_add\_\_}) must be defined at the class scope.
         Therefore during \code{refactoring}, all operators defined at the global scope, but that could be defined at the class scope, are moved as a method of the class.
	\item[\code{Cleaning}] The \code{cleaning} operation removes useless nodes and edges in the ASG.
	     A library often depends on external libraries and headers.
         There are therefore a lot of \proglang{C++} components, defined by external headers, that are not instantiated and used by the \proglang{C++} code of the actual library.
         First, in order to remove only these useless nodes, all nodes are marked as \textit{removable}.
         Then, nodes defined by the internal library are marked as \textit{non-removable}.
         Recursively, all dependencies of nodes marked as non-removable are marked as \textit{non-removable}.
         Finally, all nodes still marked as \textit{removable} are removed from the ASG.
         Some \proglang{C++} libraries, such as \pkg{armadillo} \citep{San10}, provide one self-contained header that only includes all library headers.
         In such cases all \proglang{C++} components will be marked as external dependency and the \code{clean} parameter of the \code{default} \code{controller} should be set to \code{False}.
         Otherwise, without any instruction, all \proglang{C++} components would be removed.
\end{description}

As soon as a \proglang{C++} library does not respect the recommended guidelines of \pkg{AutoWIG} , the end-user has to implement a \code{controller}.
As stated above, this \code{controller} will ensure that code generated by the \code{Generate} step is flawless.
This step mostly consists in the addition of information concerning memory management, undefined symbols and erroneous methods of class template specializations or undesired \proglang{C++} components in \proglang{Python} (see Section~\ref{subsec:results:clanglite} for an example).

\paragraph{Generators}
Currently, \pkg{AutoWIG} provides one \code{generator} for wrapping \proglang{C++} libraries using the \pkg{Boost.Python} library.
\pkg{AutoWIG} could generate wrappers in the  \proglang{C} interface that extend the \proglang{Python} interpreter, but this low-level approach does not provide the abstraction needed to consider the requirements presented in Section~\ref{sec:requirements}.
Thus, there are many benefits in using one of the semi-automatic approaches (e.g., \pkg{Boost.Python}, \pkg{SWIG}) within wrappers code.

In particular, \pkg{AutoWIG} uses the \pkg{Boost.Python} library  to propose:
\begin{itemize}
    \item An automatic \proglang{Python} documentation using  \proglang{C++} documentation since documentation strings can be injected directly in wrappers.
    \item A consistent adaptation of \proglang{C++} patterns to \proglang{Python} thanks to globally registered type coercions, possible manipulation of \proglang{Python} objects in \proglang{C++}, and an efficient overloaded function handling.
	\item A consistent memory management thanks to the definition of call policies which can be used to handle references and pointers.
    \item An automatic translation of \proglang{C++} exceptions into \proglang{Python} errors using \proglang{C++} exceptions handling and conversion into \proglang{Python} errors.
	\item An automatic management of dependencies thanks to automatic cross-module type conversions.
\end{itemize}
The \code{boost\_python} \code{generator} was therefore designed to generate \pkg{Boost.Python} wrappers:
\begin{verbatim}
def boost_python_generator(asg, nodes, module='./module.cpp',
                           decorator=None, closure=True,
                           prefix='wrapper_'):
    ...
    return wrappers
\end{verbatim}
\pkg{Boost.Python} uses extensively \proglang{C++} class templates.
However, class templates may use a huge amount of memory that can entail compilation problems.
To avoid this kind of problems, our implementation mainly consists in dispatching wrapper code for \proglang{C++} components (\code{nodes}) into different files:
\begin{description}
    \item[Module file]
    	 A module file is created in the ASG and named according to the \code{module} parameter.
         This module file is associated with multiple export files (see below).
         Its content corresponds to the inclusion of wrappers defined in their associated export files within a \code{BOOST\_PYTHON\_MODULE} block.
         The compilation of this file produces a \proglang{Python} library containing all the \proglang{C++} wrapped components.
         This library has the same basename as the module file prefixed by an underscore.
    \item[Export files]
         Export files are created in the ASG within the same directory as the module file.
         Their content declares \pkg{Boost.Python} wrappers for associated \proglang{C++} components.
         The export file of a \proglang{C++} component is named by the concatenation of its \code{prefix} parameter and an unique identifier (created from the global name hash).
         As a consequence, \pkg{AutoWIG} creates as many files as namespaces, enumerators, variables, bunch of overloaded functions and classes given in the \code{nodes} parameter.
         Note that enumerators, fields and methods wrappers are included in their parent scope export file.
         Moreover, in order to prevent name collision in \proglang{Python}, \proglang{C++} components are wrapped in \proglang{Python} modules corresponding to their \proglang{C++} scope.
    \item[Decorator file]
    	 A decorator file, named according to the \code{decorator} parameter, is created in the ASG (if \code{decorator} is not set to \code{None}).
         The \pkg{Boost.Python} library does not provide a way to wrap aliases.
         Moreover, for serialization purposes, member (i.e., class scoped declarations) classes or enumerations must not be wrapped as class member but as module member.
         The decorator file therefore contains \proglang{Python} code to define aliases or produce member aliases for member classes or enumerations.
         Note that, in some cases, programmers want to decorate the \proglang{C++} like interface into a more common \proglang{Python} interface.
         For this purpose, the decorator contains lists grouping for a template class all its instantiations.
         This allows to select easily all these instantiations in order to decorate them in the same way.
\end{description}
The code written in each of these files is generated using the \pkg{Mako} templating engine \citep{Bay12}.
Template engines are classically used in Web frameworks to generate dynamic \proglang{HTML} pages.
In our case, we use a template language to generate automatically \proglang{C++} wrapper code from patterns found in the ASG.
Changing code generation would require only to change the template code.
In order to provide a modular wrapper generation, templates must be encapsulated into classes.
Class selection for previous files is governed by plugin managers (see Table~\ref{table:architecture:generator:boost_python}).

If the parameter \code{closure} is set to \code{True}, all the dependencies of the input \proglang{C++} components (\code{nodes}) are also wrapped if they are not explicitly marked as non-exportable.
To mark a node as non-exportable, its \code{boost\_python\_export} property has to be set to \code{False} (see Section~\ref{subsec:results:clanglite} for an example).
Note that the \code{boost\_python} \code{generator} does not respect the contract of \code{generator} plugins since it requires \code{asg} and \code{nodes} as inputs, in place of requiring only \code{asg}.
In fact, this implementation is used in all other \code{generator} implementations that only needs to define abstract semantic graph (\code{asg}) traversals to compute \code{nodes} that will be considered as inputs of the \code{boost\_python} \code{generator}:
\begin{itemize}
	\item The \code{boost\_python\_internal} \code{generator} selects all nodes that are declared in headers marked as internal dependency headers.
    \item \code{boost\_python\_pattern} \code{generator} selects all nodes that match a regular expression denoted by the \code{pattern} parameter.
    This \code{pattern} parameter is set by default to \code{".*"}, so all nodes are considered.
\end{itemize}

\begin{table}
	\begin{center}
    	\begin{tabular}{c|c}
        	\multicolumn{2}{c}{Plugin}\\
            manager & finality \\\hline\hline
            \multirow{5}{*}{\code{boost\_python\_export}} & \multirow{5}{*}{\parbox{.3\textwidth}{Returning a class containing templates for the generation of \pkg{Boost.Python} wrappers for \proglang{C++} components.}}\\
            & \\
            & \\
            & \\
            & \\\hline
            \multirow{5}{*}{\code{boost\_python\_module}} & \multirow{5}{*}{\parbox{.3\textwidth}{Returning a class containing templates for the generation of \pkg{Boost.Python} module for \pkg{Boost.Python} wrappers.}}\\
            & \\
            & \\
            & \\
            & \\\hline
            \multirow{5}{*}{\code{boost\_python\_decorator}} & \multirow{5}{*}{\parbox{.3\textwidth}{Returning a class containing templates for the generation of \proglang{Python} code to complete \pkg{Boost.Python} wrappers.}}\\
            & \\
            & \\
            & \\
            & \\\hline
        \end{tabular}
        \caption{Plugin managers to control the \code{boost\_python} \code{generator}.
        	     $3$ plugin managers are used in the \code{boost\_python} \code{generator}.
                 This enable the choice of \pkg{Mako} templates \citep{Bay12} to compute the content of wrappers.
                 The generation of wrappers is therefore customizable.}
        \label{table:architecture:generator:boost_python}
    \end{center}
\end{table}

\section[C++ coding guidelines]{\proglang{C++} coding guidelines}
\label{sec:guidelines}

Considering the requirements presented in Section~\ref{sec:requirements}, we recommend to use the following guidelines in order to benefit from the most automated wrapping procedure.

\paragraph{Parse self-contained headers}
An \pkg{AutoWIG} \code{parser} requires self-contained headers.
In other words, a header should have header guards, should include all other headers it needs, and should not require any particular symbols to be defined.
Any non self-contained headers, should not be given to a \code{parser} but can nevertheless be considered during parsing using relevant search path flags (given by the \code{-I} option).

\paragraph{Use smart pointers}
Let us consider a \proglang{C++} template function declaration that returns a pointer,
\begin{verbatim}
	template<class T> T* ambiguous_function();
\end{verbatim}
There is \emph{a priori} no way to know whether the pointer should be deleted or not by the caller.
\pkg{Boost} and \pkg{STL} (Standard Template Library) libraries have introduced smart pointers as a design pattern to ensure correct memory management.
Smart pointers (i.e., \code{unique\_ptr}, \code{shared\_ptr} and \code{weak\_ptr}) define how to manage the memory of a pointer, take the responsibility to delete the pointer, and thus remove these \proglang{C++} ambiguities.
In the following example,
\begin{verbatim}
	template<class T> std::unique_ptr< T > unambiguous_function();
\end{verbatim}
the usage of \code{std::unique\_ptr} explicits the fact that the caller takes ownership of the result, and the \proglang{C++} runtime ensures that the memory for \code{T*} will be reclaimed automatically.
By default, \pkg{AutoWIG} considers that any raw pointer should not be deleted by the caller.
If this is not the case, \pkg{Boost.Python} call policies can be set to ensure proper memory management.


\paragraph{Use \proglang{C++} \pkg{STL} containers}
In \proglang{C++}, containers can be expressed as \proglang{C} arrays (e.g., \code{double array[10];}) or pointers to arrays (\code{double* ptrarray = array;}).
However, \proglang{C++} components (e.g., variables, functions) that are using \proglang{C} arrays or pointers to arrays are not wrapped by the \code{boost\_python} \code{generator} due to ambiguity.
In these cases, we recommend to use \proglang{C++} arrays (e.g., \code{std::array< double, 10 >}) or dynamic arrays (e.g., \code{std::vector< double >}), which can be effectively wrapped using the \code{boost\-\_python} \code{generator}.

\paragraph{Derive from \code{std::exception}}
In \proglang{C++}, exceptions provide a way to react to exceptional circumstances in programs, like runtime errors, by transferring control to special functions called handlers.
The \pkg{C++ standard library} provides a base class -- \code{std::exception} defined in the \code{<exception>} header -- especially designed to declare objects to be thrown as exceptions.
By default, for a \proglang{Python} interfaced \proglang{C++} library, \pkg{Boost.Python} translates a \proglang{C++} exception thrown by wrapped functions or module into a \proglang{Python} \code{RuntimeError}.
To produce better error messages, \pkg{AutoWIG} ensures that any exception derived from the \code{std::exception} class is correctly translated (i.e., the error raised has the same class name and content).

\paragraph{Pay attention to \code{static} and \code{const} overloading}
Let us consider the header presented in Figure~\ref{fig:guidelines:overload}.
We here assume that the library has been wrapped using \pkg{AutoWIG} in an \pkg{basic} \proglang{Python} package.
\begin{verbatim}
>>> from basic import Overload
>>> overload = Overload()
\end{verbatim}
\proglang{Python} is not designed for function overloading but \pkg{Boost.Python} provides some meta-program\-ming mechanisms in order to perform dispatching and therefore enable function overloading in \proglang{Python}.
Yet, considering \code{static} and \code{const} specifiers, few problems can arise:
\begin{itemize}
	\item Overloading a function with \code{static} renders all overloaded methods as \code{static} methods.
          If this entails strange usage of methods that are actually not \code{static}, it remains possible to call all overloaded methods.
\begin{verbatim}
>>> overload.staticness(overload)
non-static
>>> Overload.staticness(overload, 0)
static
\end{verbatim}
          Yet, if \code{static} overload has for first parameter an instance, a reference or a pointer to its parent class and all following parameters corresponding to another non-\code{static} overload, the non-\code{static} method will not be callable in the \proglang{Python} interpreter.
\begin{verbatim}
>>> Overload.staticness(overload, 0)
static
>>> overload.staticness(overload, 0)
static
\end{verbatim}

    \item Overloading a function with \code{const} hides the previous one written in the header.
\begin{verbatim}
>>> overload.constness()
const
>>> overload.nonconstness()
non-const
\end{verbatim}
		  This can have serious side effects on the library usage.
          We therefore recommend to specify in the \code{controller} implementation which overload must not be considered, or to design headers considering this rule.
\end{itemize}

\begin{figure}
	\begin{center}
\begin{verbatim}
/**
 * \brief This class is used to illustrate problems that can arise with
 *        overloading
 * \details At this stage mainly static (\ref ::Overload::staticness) 
 *          and const (\ref ::Overload::constness or 
 *          \ref ::Overload::nonconstness) overloading are reported as
 *          problematic.
 * \note The documentation is also used for illustrating the Doxygen to
 *       Sphinx conversions
 * \todo Any problem concerning method overloading should be added in this
 *       class
 * */
struct Overload 
{
    Overload();
 
    /// \brief This method print "static" in the standard C output stream
    void staticness();
    
    /// \brief  This method print "static" in the standard C output stream
    void staticness(const unsigned int value); 

    /// \brief This method print "non-static" in the standard C output
    ///        stream
    static void staticness(const Overload& overload,
                           const unsigned int value);

    /// \brief print "non-const" in the standard C output stream
    void constness();

    /// \brief print "const" in the standard C output stream
    void constness() const;

    /// \brief print "const" in the standard C output stream    
    void nonconstness() const;

    /// \brief print "non-const" in the standard C output stream
    void nonconstness();
};
\end{verbatim}
	\caption{\label{fig:guidelines:overload} A basic header used for illustrating overloading problems.
            The method \code{void staticness(const unsigned int value)} (resp. \code{void constness()} or \code{void nonconstness() const}) can be wrapped but as soon as \code{static void staticness(const Overload\& overload, const unsigned int value)} (resp. \code{void constness() const} or \code{void nonconstness()}) is also wrapped, it will not be callable in the \proglang{Python} interpreter.}
	\end{center}
\end{figure}

\paragraph{Use namespaces}
Namespaces prevent name conflicts in large projects.
Symbols declared inside a namespace block are placed in a named scope that prevents them from being mistaken for identically-named symbols in other scopes.
The usage of a base namespace for each \proglang{C++} library (e.g., \code{std}, \code{boost}) is highly recommended since it ease code introspection with \pkg{AutoWIG}.

\paragraph{Document with \pkg{Doxygen} and \pkg{Sphinx}}
For \proglang{C++} documentation, \pkg{Doxygen} \citep{Hee08} is one of the most standard tool for generating formatted, browsable, and printable documentation from annotated sources.
Its equivalent for \proglang{Python} is \pkg{Sphinx} \citep{Bra09}.
Writing and verifying documentation is a fastidious task, and the redundancy between \proglang{C++} and \proglang{Python} wrapped components must be limited.
As illustrated below, \pkg{AutoWIG} parses the \pkg{Doxygen} documentation in the \proglang{C++} code source (see Figure~\ref{fig:guidelines:overload}) and formats it into a \pkg{Sphinx} documentation. This documentation string is then injected into the \proglang{Python} components.
\begin{verbatim}
>>> help(overload)
...
This class is used to illustrate problems that can arise with
overloading

At this stage mainly static
(:py:meth:`test.overload._bar.Overload.staticness`) and
const (:py:meth:`test.overload._bar.Overload.constness` or
:py:meth:`test.overload._bar.Overload.nonconstness`)
overloading are reported as problematic.

.. note::

    The documentation is also used for illustrating the Doxygen to Sphinx
    conversions

.. todo::

    Any problem concerning method overloading should be added in this class
...
\end{verbatim}

\section{Results}
\label{sec:results}

In the following section, we present some examples using \pkg{AutoWIG} in order to emphasize particular aspects of the wrapping process.
Therefore, most of the presented examples are truncated or modified for the sake of clarity and simplicity.
Nevertheless, these examples are all fully available and reproducible on a notebook server (see Section~\ref{subsec:discussion:installation} and supplementary materials for details).

\subsection{Wrapping a basic library}
\label{subsec:results:simple}

We here aim at presenting the interactive wrapping workflow.
For the sake of simplicity, we consider a basic example of \proglang{C++} library (see header presented in Figure~\ref{fig:results:simple}).

\begin{figure}
	\begin{center}
\begin{verbatim}
#include <exception>

struct ProbabilityError : std::exception
{
    /// \brief Compute the exception content
    /// \returns The message "a probability must be in the interval [0,1]"
    virtual const char* what() const noexcept; 
};

struct BinomialDistribution
{
    BinomialDistribution(const unsigned int n, const double pi);
    BinomialDistribution(const BinomialDistribution& binomial);
    ~BinomialDistribution();
    
    //! \brief Compute the probability of a value
    //! \details The probability is given by the flowwing formula \cite{JK
    //!          
    //!          \f{equation*}{
    //!                 P\left(X = x\right) = \begin{cases}
    //!                                            \binon{n}{x} \pi^x \lef
    //!                                             0 & \mbox{otherwise}
    //!                                       \end{cases}.
    //!          \f}
    //! \returns The probability
    double pmf(const unsigned int value) const;
        
    double get_pi() const;
    
    /** 
     * \param pi New probability value
     * \warning The probability value must be in the interval  \f$\left[0,
     * \throws \ref ::ProbabilityError If the new probability value is not 
     *         in the interval \f$\left[0,1\right]\f$ */
    void set_pi(const double pi);
   
    unsigned int n;
};
\end{verbatim}
	\caption{\label{fig:results:simple} A header for a basic library.
            This basic \proglang{C++} library implements probability mass function computation for binomial distributions (\code{BinomialDistribution::pmf}).
            If an user try to set the probability parameter of the binomial distribution (\code{BinomialDistribution::\_pi}) to values outside the interval $[0,1]$, a \code{ProbabilityError} exception is thrown.}
	\end{center}
\end{figure}

First, import \pkg{AutoWIG}.
\begin{verbatim}
>>> import autowig
\end{verbatim}
Assuming that the header is located at \code{'./basic/binomial.h'}, we parse it with relevant compilation flags.
\begin{verbatim}
>>> asg = autowig.AbstractSemanticGraph()
>>> asg = autowig.parser(asg, ['./basic/binomial.h'],
...                           ['-x', 'c++', '-std=c++11'])
\end{verbatim}
Since most of \pkg{AutoWIG} guidelines are respected, the \code{default} \code{controller} implementation is suitable.
\begin{verbatim}
>>> autowig.controller.plugin = 'default'
>>> asg = autowig.controller(asg)
\end{verbatim}
In order to wrap the library we need to select the \code{boost\_python\_internal} \code{generator} implementation.
\begin{verbatim}
>>> autowig.generator.plugin = 'boost_python_internal'
\end{verbatim}
The \pkg{Boost.Python} module (resp. decorator) name chosen is \code{'./basic/module.cpp'} (resp. \code{'./basic/\_module.py'}).
\begin{verbatim}
>>> wrappers = autowig.generator(asg, module = './basic/module.cpp',
...                                   decorator = './basic/_module.py')
\end{verbatim}
The wrappers are only generated in-memory.
We therefore need to write them on the disk to complete the process.
\begin{verbatim}
>>> wrappers.write()
\end{verbatim}
Once the wrappers are written on disk, we need to compile and install the \proglang{Python} bindings.
Finally, we can use the \proglang{C++} library in the \proglang{Python} interpreter
\verbatiminput{binomial.py}

\subsection{Wrapping a subset of a very large library}
\label{subsec:results:clanglite}

Sometimes, for a very large library, only a subset of available \proglang{C++} components is useful for end-users.
Wrapping such libraries therefore requires \pkg{AutoWIG} to be able to consider only a subset of the \proglang{C++} components during the \code{Generate} step.
The \pkg{Clang} library is a complete \proglang{C}/\proglang{C++} compiler.
\pkg{Clang} is a great tool, but its stable \proglang{Python} interface (i.e., \pkg{libclang}) is lacking some useful features that are needed by \pkg{AutoWIG}.
In particular, class template specializations are not available in the abstract syntax tree.
Fortunately, most of the classes that would be needed during the traversal of the \proglang{C++} abstract syntax tree are not template specializations.
We therefore proposed to bootstrap the \pkg{Clang} \proglang{Python} bindings using the \code{libclang} \code{parser} of \pkg{AutoWIG}.
This new \pkg{Clang} \proglang{Python} interface is called \pkg{ClangLite} and is able to parse class template specializations.
As for \pkg{libclang}, this interface is proposed only for a subset of the \pkg{Clang} library sufficient enough for proposing the new \code{clanglite} \code{parser}.

In order to wrap a library subset, the user need to define a \code{controller} implementation that specifies which \proglang{C++} components will be considered during the \code{Generate} step.
The \code{controller} implemented is the following:
\begin{verbatim}
def clanglite_controller(asg):
    ...
    for node in asg.classes():
        node.boost_python_export = False
    for node in asg.enumerations():
        node.boost_python_export = False
    ...
    subset = []
    classes = [asg['class ::clang::Type'], asg['class ::clang::Decl']]
    subset += classes
    subset += classes[0].subclasses(recursive = True)
    subset += classes[1].subclasses(recursive = True)
    subset.append(asg['class ::llvm::StringRef'])
    ...
    for node in subset:
        node.boost_python_export = True
    ...
    return asg
\end{verbatim}
This \code{clanglite} \code{controller} principally consists in:
\begin{enumerate}
	\item Considering all user-defined types as non-exportable.
          This is done by setting the \code{boost\_python\_export} property of classes and enumerations to \code{False} (lines 3--6).
    \item Considering a subset of all user-defined types as exportable.
          This is done by first selecting the \proglang{C++} components of interest (\code{subset}) using code introspection (lines 8--13).
          Then, the \code{boost\_python\_export} property of all subset components is set to \code{True} (lines 15--16).
\end{enumerate}

Assuming that the \code{asg} already contains all \proglang{C++} components from the \pkg{Clang} library and that the \code{clanglite\_controller} has been defined in the \code{Python} interpreter.
We need to register the \code{clanglite\_controller} as a \code{controller} implementation and then to select it.
\begin{verbatim}
>>> autowig.controller['clanglite'] = clanglite_controller
>>> autowig.controller.plugin = 'clanglite'
\end{verbatim}
After the generation and compilation of wrappers (using the same procedure as the one described in Section~\ref{subsec:results:simple}), it enabled us to propose a new \code{parser} implementation called \code{clanglite}.
This has been done by writing \code{Python} code responsible for the traversal of the AST and the completion of an existing ASG.
Contrarily to the \code{libclang} \code{parser} the AST traversed by the \code{clanglite} \code{parser} contains template classes and their specializations.
This \code{parser} is therefore more efficient and is selected by default in \pkg{AutoWIG}, as soon as the \pkg{ClangLite} bindings are installed.


\subsection{Wrapping a template library}
\label{subsec:results:pystl}

A template library is a library where there are only template classes that can be instantiated.
Wrapping such libraries therefore requires \pkg{AutoWIG} to be able to consider various \proglang{C++} template classes instantiations during the \code{Parse} step.
The \pkg{Standard Template Library (STL)} library \citep{PLMS00} is a \proglang{C++} library that provides a set of common \proglang{C++} template classes such as containers and associative arrays.
These classes can be used with any built-in or user-defined type that supports some elementary operations (e.g., copying, assignment).
It is divided in four components called algorithms, containers, functional and iterators.
\pkg{STL} containers (e.g., \code{std::vector}, \code{std::set}) are used in many \proglang{C++} libraries.
In such a case, it does not seem relevant that every wrapped \proglang{C++} library contains wrappers for usual \pkg{STL} containers (e.g., \code{std::vector< double >}, \code{std::set< int >}).
We therefore proposed \proglang{Python} bindings for some sequence containers (e.g., \code{vector} of the \code{std} namespace) and associative containers (e.g., \code{set}, \code{unordered\_set} of the \code{std} namespace).
These template instantiations are done for various \proglang{C++} fundamental types (e.g., \code{int} , \code{unsigned long int}, \code{double}) and the \code{string} of the \code{std} namespace).
For ordered associative containers only the \code{std::less} comparator was used.

In order to wrap a template library, the user needs to write headers containing aliases for desired template class instantiations:
\begin{verbatim}
#include <vector>
#include <string>
...
typedef std::vector< unsigned long int > VectorUnsignedLongInt;
typedef std::vector< int > VectorInt;
typedef std::vector< double > VectorDouble;
typedef std::vector< std::string > VectorString;
...
\end{verbatim}
After the generation and compilation of wrappers (using the same procedure as the one described in Section~\ref{subsec:results:simple}), the user can hereafter use \proglang{C++} containers in the \proglang{Python} interpreter.
\begin{verbatim}
>>> import stl
>>> v = stl.VectorInt()
>>> v.push_back(-1)
>>> v.push_back(0)
>>> v.push_back(1)
\end{verbatim}
Note that in order to have a functional \proglang{Python} package, some methods can be dynamically added to wrapped classes within modules.
For instance, in the \code{stl/vector.py} module:
\begin{itemize}
	\item The \code{\_\_iter\_\_} method that enables iterations over a wrapped vector and its conversion to \proglang{Python} list is added to all \code{std::vector} class template instantiations wrapped.
	\begin{verbatim}
>>> list(v)
[-1, 0, 1]
	\end{verbatim}
    \item The \code{\_\_str\_\_} and \code{\_\_repr\_\_} methods that enable representations in the \proglang{Python} interpreter of vectors are added to all \code{std::vector} class template instantiations wrapped.
    \begin{verbatim}
>>> v
(-1, 0, 1)
	\end{verbatim}
\end{itemize}
Moreover, the \code{stl/\_\_init\_\_.py} module imports all \proglang{Python} objects of the \code{stl/\_\_stl.so} library at its root to simplify class names (e.g., \code{stl.VectorInt} instead of \code{stl.\_\_stl.VectorInt}).

Some additional features are automatically added in \pkg{AutoWIG} wrappers.
For example, for functions returning non-constant references (e.g.,  \texttt{int\& operator[] (size\_type pos);} of the \code{std::vector< int >} instantiation), an additional wrapping is done using the following decorator.
\begin{verbatim}
namespace autowig
{
    method_decorator_64cf5286bbd05b06844aa126bb40d4c3(
        class std::vector< int, std::allocator<int> > & instance,
        unsigned long int  param_in_0, const int  & param_out)
    { instance.operator[](param_in_0) = param_out; }
}
\end{verbatim}
This decorator is then dealt as an overloaded method in wrappers.
In this particular example, it enables to define \code{\_\_getitem\_\_} and \code{\_\_setitem\_\_} methods in the \code{stl/vector.py} module.
\begin{verbatim}
>>> v[0]
-1
>>> v[0] = -2
>>> v[0]
-2
\end{verbatim}
If this decorator is not written, there is no way to use the \code{\_\_setitem\_\_} method in \proglang{Python}.
Moreover, since \proglang{Python} users are more familiar with \code{Python} containers, each method taking a \proglang{C++} container constant reference or copy as parameter try to convert automatically \proglang{Python} objects into the corresponding \proglang{C++} container.
Therefore, as illustrated below, \proglang{Python} list of integers are automatically converted into \proglang{C++} vectors of integers.
\begin{verbatim}
>>> stl.VectorInt([0, 1])
(0, 1)
\end{verbatim}

\subsection{Wrapping dependent libraries}
\label{subsec:results:statistic}

\begin{figure}
	\centering
	\subfloat[]{\includegraphics[width=.4\textwidth]{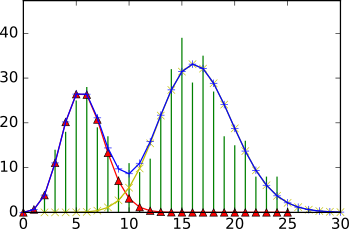}
	\label{fig:mm_assessment:d}}
    \hfil
	\subfloat[]{\includegraphics[width=.4\textwidth]{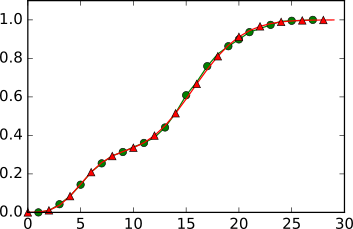}
	\label{fig:mm_assessment:d}}\\
	\subfloat[]{\includegraphics[width=.4\textwidth]{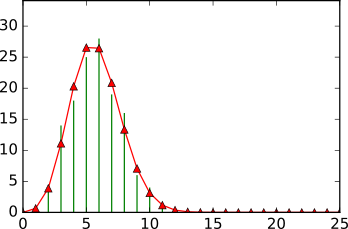}
	\label{fig:mm_assessment:d}}
    \hfil
	\subfloat[]{\includegraphics[width=.4\textwidth]{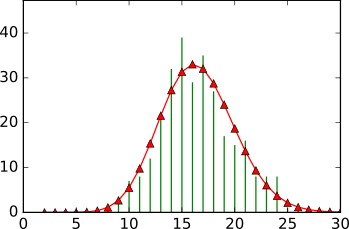}
	\label{fig:mm_assessment:d}}
    \caption[Mixture model]{Visualizations proposed by the \pkg{StatTool} \proglang{Python} bindings for mixture model quality assessment.\label{fig:mm_assessment}
            (a) The data frequency distribution is represented in green.
                The theoretical frequency distribution of the fitted mixture model with $2$ components is represented in blue.
            (b) The empirical cumulative distribution function is represented in green.
                The cumulative distribution function of the fitted mixture model with $2$ components is represented in red.
            (c) (resp. (d)) The empirical probability mass function for the data subset corresponding to the first (resp. second) component is represented in green.
                The probability mass function of the first (resp. second) component of the fitted mixture model with $2$ components is represented in red.
            }
\end{figure}

\begin{figure}
	\centering
	\subfloat[]{\includegraphics[width=.4\textwidth]{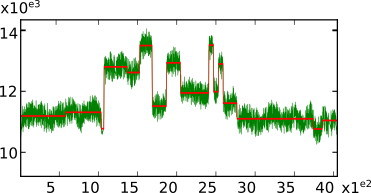}
	\label{fig:segmentation}}
    \hfil
	\subfloat[]{\includegraphics[width=.4\textwidth]{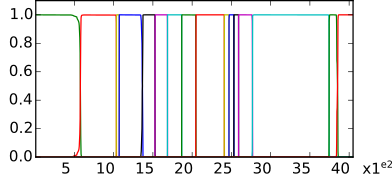}
	\label{fig:probabilities}}
    \caption[Segmentation]{Visualizations proposed by the \pkg{SequenceAnalysis} \proglang{Python} bindings for segmentation quality assessment.\label{fig:seg_assessment}
            (a) In green the nuclear-magnetic response of underground rocks is represented in function of the depth.
            Segment means are represented by the red piecewise constant function.
            (b) Posterior segment probabilities.}
\end{figure}

\pkg{StructureAnalysis} is a set of libraries including statistical models for the analysis of structured data (mainly sequences and tree-structured data):
\begin{itemize}
    \item \pkg{StatTool} is a library containing classes for the parametric  modeling of univariate and multivariate data (see Figure~\ref{fig:mm_assessment}).
    \item \pkg{SequenceAnalysis} is a library containing statistical functions and classes for markovian models (e.g., hidden variable-order Markov and hidden semi-Markov models) and multiple change-point models for sequences  (see Figure~\ref{fig:seg_assessment}).
    	The \pkg{SequenceAnalysis} library depends on the \pkg{StatTool} library.
\end{itemize}
These libraries have been extensively used for the identification and characterization of developmental patterns in plants from the tissular to the whole plant scale.
Previously interfaced with \proglang{AML} (a home-made, domain-specific programming language), some work has been done to switch to \proglang{Python}.
Nevertheless, the complexity of writing wrappers with \pkg{Boost.Python} limited the number of available components in \proglang{Python} in comparison to \proglang{AML}.
One advantage of having a statistical library written in \proglang{C++} available in \proglang{Python} is that developers can benefit from all other \proglang{Python} packages.
As illustrated in Figures~\ref{fig:mm_assessment}--~\ref{fig:seg_assessment}, this is particularly useful for providing visualizations for statistical model assessment using -- for example -- the \pkg{Matplotlib} \citep{Hun07} \proglang{Python} package.

\paragraph[The StatTool library]{The \pkg{StatTool} library}
In order to wrap a \proglang{C++} library, that will be used as a dependency by other libraries, the user needs to save the ASG resulting from the wrapping process.
In the \pkg{StatTool} case, we first generate the wrappers (using the same procedure as the one described in Section~\ref{subsec:results:simple}).
Then, we use the \pkg{pickle} \proglang{Python} package for serializing the \pkg{StatTool} ASG in the \code{'ASG.pkl'} file.
\begin{verbatim}
>>> import pickle
>>> with open('ASG.pkl', 'w') as f:
...   pickle.dump(asg, f)
\end{verbatim}

After the compilation of the wrapper, the user can hereafter use mixture models in the \proglang{Python} interpreter.
For instance, we considered an example concerning the identification of preformed and neoformed parts in plants.
\begin{verbatim}
>>> from structure_analysis import stat_tool
>>> his = stat_tool.Histogram("meri.his")
\end{verbatim}
The data  (\texttt{his}) consists of the number of elongated organs of $424$ shoots of wild cherry tree (\emph{Prunus avium}).
These shoots were sampled in different architectural positions (from the trunk to peripheral positions of the trees) and were representative of the full range of growth potential.
The proximal part of a shoot always consists of preformed organs (i.e., organs contained in the winter bud).
This preformed part may be followed by a neoformed part consisting of organs differentiated and elongated during the current growing season.
We estimated mixture of parametric discrete distributions on the basis of this data.
The number of components ($2$) was selected between $1$ and $4$ using the bayesian information criterion.
\begin{verbatim}
>>> mixt = stat_tool.MixtureEstimation(meri, 1, 4, "BINOMIAL")
...
1 distribution   2 * log-likelihood: -2735.4   3 free parameters
                 2 * penalyzed log-likelihood (BIC): -2753.54
                 weight: 1.17894e-28

2 distributions   2 * log-likelihood: -2587.18   7 free parameters
                  2 * penalyzed log-likelihood (BIC): -2624.93
                  weight: 0.99791

3 distributions   2 * log-likelihood: -2581.43   11 free parameters
                  2 * penalyzed log-likelihood (BIC): -2637.27
                  weight: 0.00208662

4 distributions   2 * log-likelihood: -2581.51   15 free parameters
                  2 * penalyzed log-likelihood (BIC): -2649.93
                  weight: 3.73165e-06

\end{verbatim}
Further investigations can be performed in order to asses the quality of the $2$ component mixture model.
For instance, we considered here the visualization of various probability functions.
\begin{verbatim}
>>> mixt.plot()
\end{verbatim}
As illustrated on Figure~\ref{fig:mm_assessment} the data are well fitted by the mixture model and:
\begin{itemize}
	\item The first component corresponds to entirely preformed shoots.
    \item The second component to mixed shoots consisting of a preformed part followed by a neoformed part.
\end{itemize}

\paragraph[The StructureAnalysis library]{The \pkg{StructureAnalysis} library}
In order to wrap a \proglang{C++} library that has dependencies, the user needs to combine the ASGs resulting from the wrapping of its dependencies before performing its own wrapping.
In the \pkg{SequenceAnalysis} case, we construct first an empty ASG.
\begin{verbatim}
>>> asg = AbstractSemanticGraph()
\end{verbatim}
Then, we use the \pkg{pickle} \proglang{Python} package for de-serializing the \pkg{StatTool} ASG (assumed to be serialized in the \code{'../stat\_tool/ASG.pkl'} file) and merge it with the current ASG.
\begin{verbatim}
>>> import pickle
>>> with open('../stat_tool/ASG.pkl', 'r') as f:
...   asg.merge(pickle.load(f))
\end{verbatim}

After the generation and compilation of wrappers (using the same procedure as the one described in Section~\ref{subsec:results:simple}), the user can hereafter use multiple change-point models \citep[see][for applications of multiple change-point models]{GCHLM07,LGMYB15} in the \proglang{Python} interpreter.
Multiple change-point models are used to delimit segments within sequences, for which the characteristics of variables (or vectors in the multivariate case) are homogeneous within each segment while differing markedly from one segment to another (e.g., piecewise constant mean and variance for a Gaussian change in the mean and variance model).
For instance, we considered the classic example of well-log data \citep{Gue13,Gue15a,Gue15b}.
\begin{verbatim}
>>> from structure_analysis import sequence_analysis
>>> seq = sequence_analysis.Sequences("well_log_filtered_indexed.seq")
\end{verbatim}
The data (\texttt{seq}) consist of $4050$ measurements of the nuclear-magnetic response of underground rocks.
The data were obtained by lowering a probe into a bore-hole.
Measurements were taken at discrete time points by the probe as it was lowered through the hole.
The underlying signal is roughly piecewise constant, with each constant segment relating to a single rock type that has constant physical properties.
The change points in the signal occur each time a new rock type is encountered.
Outliers were removed before the data were analyzed.
We estimated Gaussian change in the mean and variance models on the basis of the well-log filtered data.
The number of segments ($16$) was selected using the slope heuristic \citep{Gue15b} with a slope estimated using log-likelihood of overparametrized models ranging from $30$ up to $80$ change points.
\begin{verbatim}
>>> seq.segmentation(0, 80, "Gaussian", min_nb_segment=30)
...
2 * log-likelihood: -68645.9
change points: 578, 1035, 1071, 1369, 1527, 1686, 1867, 2048, 2410, 2470,
               2532, 2592, 2769, 3745, 3856
segment sample size: 535, 439, 15, 277, 146, 151, 164, 170, 344, 55, 57, 58,
                     169, 926, 106, 152
segment mean, standard deviation: 111907 2241.11 | 113095 2313.87
                                  107735 1635.75 | 128010 2229.5 
                                  126154 2106.38 | 134990 2311.48
                                  115124 2037.82 | 129330 2269.02
                                  119454 2085.75 | 135167 1990.3 
                                  119852 2196.84 | 128968 1797.23
                                  116114 2117.15 | 110981 2284.67
                                  107661 2098.15 | 110430 2328.86
...
\end{verbatim}
Further investigations can be performed in order to asses the non-ambiguous character of the segmentation into $16$ segments.
For instance, we considered here the visualization of segment profiles \citep[][see Figure~\ref{fig:seg_assessment}]{Gue13, Gue15a}.
\begin{verbatim}
>>> prf = seq.segment_profile(1, 16, "Gaussian")
>>> prf.plot()
\end{verbatim}


\section{Discussion}
\label{sec:discussion}

\subsection{Related work}

\proglang{Python} and \proglang{R} are interpreted languages implemented in \proglang{C}.
Like many other scripting languages, they provide a \proglang{C} API (i.e., Application Programming Interface) to allow foreign libraries implemented in \proglang{C} or in a language compatible with \proglang{C} (e.g., \proglang{C++} or \proglang{Fortran}) to extend the language.
This design feature has been a key element for the adoption of the \proglang{Python} language as a glue language, by providing efficient standard libraries implemented in compiled languages.
This \proglang{C} API is designed to be stable but low-level.
It does not provide support for object-oriented languages, and every type and function have to be \textit{manually} wrapped.
Note that if this approach is only efficient for exposing few functions and objects for developers, it is also at the basis of all other wrapper tools that generate \proglang{C} API code.

Several \textit{semi-automatic} solutions (e.g., \pkg{Cython}, \pkg{SWIG} and \pkg{Boost.Python}) have been proposed to simplify and ease the process of wrapping large \proglang{C++} libraries.
\pkg{SWIG} \citep{Bea03,Bea09} implements its own compiler that simplifies the process of wrapping large \proglang{C} and \proglang{C++} libraries into a large number of different languages, and in particular \proglang{R} and \proglang{Python}.
%
%
%
%
%
While \pkg{SWIG} is capable of wrapping most of the \proglang{C++} features, it requires configuration files to include interface and conversion specifications.
%
%
%
If there is a change in the library, these configuration files may become out of date.
\pkg{Cython} \citep{BBCD+11} is another semi-automatic solution.
\pkg{Cython} both enables \proglang{Python} users to compile \proglang{Python} code to \proglang{C} for optimizing execution of scientific code, and makes it possible for developers to call \proglang{C} or \proglang{C++} code from \proglang{Python}.
\pkg{Cython} is intensively used by several \proglang{Python} scientific libraries \citep{PVGMTGBPWD+11, WSNBWYGY14} that optimized critical part of their code by writing subparts of the package in \pkg{Cython}.
It has been originally developed as part of the \pkg{Sage} project \citep{sage15} to integrate numerous packages and libraries written in \proglang{C}, \proglang{C++} and \proglang{Fortran}.
However, \pkg{Cython} requires re-declaration of every class and function to wrap a \proglang{C} or \proglang{C++} library.
Finally, \pkg{Boost.Python} \citep{AG03} and \pkg{Rcpp} \citep{EFAC+11} depend on meta-programming to provide high-level abstractions (e.g., registration of classes and inheritance, automatic conversion of registered types and classes,  management of smart pointers, \proglang{C++} object-oriented interface to \proglang{Python} objects, ...).
However, all the wrappers have to be written and keep in sync with the code of the library, and require lots of knowledge for developers.

Recently, several projects have provided \emph{automatic} solutions for wrapping existing \proglang{C++} libraries.
They mainly rely on the same kind of architecture:
\begin{itemize}
	\item A parser or compiler that extracts information about the list of \proglang{C++} functions or classes and their signatures.
    \item Strategies to convert this abstract view of the \proglang{C++} code into manual or semi-automatic wrapper tools.
    \item The generation of the \proglang{Python} or \proglang{R} bindings based on these information.
\end{itemize}
The first difficulty is to parse large \proglang{C++} code, and provide information on its structure.
For this, tools like \pkg{Doxygen} or \pkg{GCC-XML} have been used.
While \pkg{Doxygen} was first developed to automatically extract and render documentation of \proglang{C++} libraries, it provides an \proglang{XML} representation of the \proglang{C++} interface that can be used to describe functions and classes.
Later, \pkg{GCC-XML} has been developed to offer a representation of a \proglang{C++} library in \proglang{XML} using the \pkg{GCC} compiler.
This tool has been developed for one of the first automatic library, \pkg{CABLE}, used to wrap the large visualization library \pkg{VTK} \citep{SML97}.
However, maintaining such a tool is complex and \pkg{GCC-XML} does not support \proglang{C++ 11} standard.
In \pkg{AutoWIG}, we use the \pkg{LLVM}/\pkg{Clang} technologies \citep{Lat08} to have the latest version of the compiler.
\pkg{Clang} provides a full representation of the compiled library.
Among the automatic tools, \pkg{CABLE} and \pkg{WrapITK} \citep{lehmann2006wrapitk} generate \pkg{SWIG} configuration files to build the wrappers, \pkg{Py++} \citep{Yak11} generates \pkg{Boost.Python} code, and \pkg{XDress} \citep{Sco13} generates \pkg{Cython} files.
Some domain specific tools, like \pkg{Shiboken}, have also been developed to wrap their large \pkg{C++} libraries (in this case the entire \pkg{QT} libraries).
While these tools provide an excellent solution for very complex libraries, they have some limitations.
Some libraries rely on \pkg{GCC-XML} that does not support modern \proglang{C++} standard.
However, a new tool \pkg{CastXML} is currently in development.
The main tools depends on configuration files and are called as executable like \pkg{XDress} and \pkg{WrapITK}.
While they can easily be integrated in development workflow, it is not easy for developers to drive and specialize them using a scripting language.
\pkg{AutoWIG} and \pkg{Py++} provide a \proglang{Python} interface and offer introspection facilities for \proglang{C++} libraries from \proglang{Python}.
Like \pkg{Py++}, \pkg{AutoWIG} generates \pkg{Boost.Python} wrappers.
However, \pkg{Py++} depends on \pkg{GCC-XML} and requires to write a full parser and code generator in \proglang{Python}.
It allows to implement a fully automatic system for developers based on their library design pattern, but is rather complex to implement.

\subsection{Extensibility}
As stated above, the plugin architecture of \pkg{AutoWIG} enables non-intrusive extensibility.
This is of great interest when considering the addition of other source or target languages.

The addition of a target language principally consists in writing \pkg{Mako} templates \citep{Bay12}.
As an example, let consider the \proglang{R} language.
In order to be able to propose automatic \proglang{R} bindings for \proglang{C++} libraries, the templates written could be based on the \pkg{Rcpp} \citep{EFAC+11} library.
This is particularly interesting since \pkg{Rcpp} wrappers are quite similar to \pkg{Boost.Python} ones.
As a matter of fact, the implementation of a \code{r\_cpp} \code{generator} is of highest priority regarding future work.
The major difficulty encountered is the lack of some features in \pkg{Rcpp} (e.g., enumeration wrapping) and particular organization of \proglang{R} packages that must be taken into account.

The addition of a source language is more problematic since it could lead to addition of new proxy classes in the abstract semantic graph.
For example, if the addition of the \proglang{C}, \proglang{Objective C} or \proglang{Objective C++} languages should be relatively easy since it can be done using the \pkg{Clang} parser and \proglang{C++} proxy classes, the addition of the \proglang{Fortran} language requires more work.
In fact, for this purpose the \pkg{Open Fortran Parser} \citep{RS12} could be used but it would require to reimplement the transformation of an abstract syntax tree to an abstract semantic graph.
Moreover, any addition of a source language must be followed with the addition of target language generator since wrapper technologies are dependent of source languages.
For \proglang{Fortran}, once the \code{parser} is implemented, this would require in addition to implement a \code{generator} potentially based on the \pkg{F2Py} \citep{Pet09} tool.

\subsection{Toward a reference guide generator}
In its current stage, \pkg{AutoWIG} translate the \pkg{Doxygen} \citep{Hee08} documentation into a \pkg{Sphinx} one \citep{Bra09} but only incorporates it in the wrappers.
This means that both \pkg{Doxygen} and \pkg{Sphinx} tools must be used to respectively generates \proglang{C++} and \proglang{Python} reference guides.
Writing a \code{generator} that would generate \pkg{Sphinx} compatible files containing the \proglang{C++} reference guide could be of great interest since it would allow to aggregate both \proglang{C++} and \proglang{Python} documentation within the same tool.

\subsection{Installation and usage}
\label{subsec:discussion:installation}
The installation of \pkg{AutoWIG} has been tested on Linux, MacOs X and Windows with \proglang{Python} 2.7.
Nevertheless, the most effective wrapping process relies on the \pkg{ClangLite} extension, that has not yet been released for Windows.
Note that wrappers generated with \pkg{AutoWIG} do not depend on \pkg{AutoWIG} and can be built on any operating system without regarding the operating system from which the wrappers were generated.
On each of these operating system, \pkg{AutoWIG} binaries are available using the \pkg{Conda} package management system.
Note that these binaries require to be installed in a specific environment that will be used for wrapper generation but not for compiling these wrappers since conflicts can occur between \pkg{AutoWIG}'s requirements and those of the wrapped library.

Moreover, \pkg{Docker} images \citep{Mer14} can be downloaded (\url{https://hub.docker.com/r/statiskit/autowig/tags}) and examples presented herein can be replayed using the Jupyter notebook \cite{PG07}.
More information can be found on \pkg{AutoWIG} documentation (\url{http://autowig.readthedocs.io}).

\subsection{Concluding remarks}

\pkg{AutoWIG} greatly simplifies the process of incorporation of compiled libraries within scripting language interpreter.
It provides the concept of ASG as \proglang{C++} code abstraction data model.
\pkg{AutoWIG} can therefore be used for \proglang{C++} code introspection in a \proglang{Python} interpreter to discover and analyze \proglang{C++} library components.
This enabled us to propose an automatic generation of \proglang{Python} bindings for \proglang{C++} libraries respecting some guidelines.
This generation of \proglang{Python} bindings is also combined with the automatic generation of pythonic interface (e.g., use of special functions, error translation, memory management and \pkg{Sphinx} formatted documentation) using \pkg{Mako}, a template language classically used in web frameworks.
Some compilation problems led us to also to consider a tool for parsing compiler errors that is particularly useful when considering the wrapping of class template specializations.

Note that a particular attention has been payed for \pkg{AutoWIG} architecture:
\begin{itemize}
	\item It has been designed as a library.
          This choice has been made since it enables interactive wrapping of compiled libraries in the high-level scripting language, \proglang{Python}.
          This interactivity use of \pkg{AutoWIG} increases the user ability to supervise or debug the wrapping process and reduces the level of expertise required to use this software.
    \item It has been designed as a plugin-oriented architecture.
          This choice has been made for extensibility purpose to enhance the adoption of \pkg{AutoWIG} by developers by simplifying the integration process of external contribution.
          While only \proglang{C++} to \proglang{Python} bindings have been implemented, \pkg{AutoWIG} plugin architecture eases the process of source (such as \proglang{C}) or target (such as \proglang{R}) language addition.
\end{itemize}

In Section \ref{sec:results}, we demonstrated the efficiency of using \pkg{AutoWIG} to wrap large and complex \proglang{C++} libraries, such as \pkg{Clang}.
Such an approach can be used to wrap other very large scientific libraries in an automatic way and enhance their diffusion to large communities of scientists that only use high-level scripting languages such as \proglang{Python} and \proglang{R}.

\section*{Acknowledgments}

We thanks Yann Gu\'edon for providing the \pkg{StructureAnalysis} libraries, an interesting guinea-pig library set.

\section*{Supplementary materials}
\label{sec:supplementary}
\begin{description}

	 \item[AutoWIG] The source code of \pkg{AutoWIG} is available on GitHub (\url{http://github.com/StatisKit/AutoWIG}).
                    The documentation is hosted on Read The Docs (\url{http://autowig.readthedocs.io}).
                    It is distributed under the CeCILL license (\url{http://www.cecill.info/licences/Licence_CeCILL_V2.1-en.html}).
                    CeCILL license is GPL compatible.
                    Examples presented in this article and the documentation are reproducible available on \pkg{Docker} images (\url{https://hub.docker.com/r/statiskit/autowig/tags/}).


\end{description}

\section*{References}

\bibliographystyle{model1-num-names}
\bibliography{references.bib}

\end{document}